\documentstyle[prl,aps,multicol,epsf]{revtex}

\begin{document}
\draft

\title{ Magnetic Anisotropy Variations and Non-Equilibrium Tunneling
in a Cobalt Nanoparticle }
\author{M. M.~Deshmukh${}^{1}$, S. Gu\'eron${}^{2}$,
E. Bonet${}^{1}$, A. N. Pasupathy${}^{1}$, S. Kleff${}^{3,4}$,
J. von Delft${}^{3,4}$, and D. C. Ralph${}^{1}$}
\address{${}^1$ Laboratory of Atomic and Solid State Physics,
Cornell University, Ithaca, NY, 14853\\
${}^2$Laboratoire de Physique des Solides,  Universit\'e Paris
Sud, 91405 Orsay, France \\
${}^3$Institut f\"ur Theoretische Festk\"orperphysik,
Universit\"at Karlsruhe, 76128 Karlsruhe, Germany\\
${}^4$Ludwig-Maximilians-Universit\"at, Theresienstr. 37, 80333 M\"unchen,
Germany}

\date{\today}
\maketitle

\begin{abstract}
We present detailed measurements of the discrete electron-tunneling
level spectrum within nanometer-scale cobalt particles as a function
of magnetic field and gate voltage, in this way probing individual quantum
many-body eigenstates inside ferromagnetic samples.
Variations among the observed levels indicate that different quantum
states within one particle are subject to different magnetic
anisotropy energies.
Gate-voltage studies demonstrate that the low-energy tunneling
spectrum is affected dramatically by the presence of non-equilibrium
spin excitations.
\end{abstract}

\par

\begin{multicols} {2}
\narrowtext

The electronic states within ferromagnetic devices are understood
surprisingly poorly.  For example, several types of experiments have
been done to characterize spin polarization near the Fermi level
\cite{meservey,ono,pasupathy,soulen,deteresa}, but different
techniques give different results.  Part of the difficulty is that
these experiments average over large numbers of states, and different
types of experiments effectively take differently weighted averages.
Recently, we have developed a
tunneling technique that can resolve the individual states within
nm-scale samples of ferromagnets \cite{gueron}.  This method has
helped motivate new descriptions of ferromagnetism that go beyond
mean-field Stoner models \cite{canali,kleff}.   However, the first
experiments left many open questions, particularly concerning the
proper description of anisotropy energies, and whether the tunneling
spectrum reflects the true electronic density of states or whether it
is modified by non-equilibrium effects.
To answer these questions, we have
introduced a gate
electrode to our devices and have also
developed lower-noise tunnel barriers, thereby allowing new types of 
measurements and greatly improving the data quality.  We present 
evidence that the
quantum states in a ferromagnetic nanoparticle are not all described
by the same anisotropy-energy function that governs the ground state
\cite{thiaville}, but the anisotropy varies from state to state.  We
show that non-equilibrium processes induced by tunneling affect the
measured spectrum, thereby explaining a larger-than anticipated
density of resonances.  The gate electrode also allows a comparison
with recent theories \cite{canali,kleff} that minority-electron
tunneling should dominate in a nanoparticle.

We will show data from both gated and non-gated devices.  In either case,
the samples contain a Co nanoparticle, separated from aluminum
electrodes by aluminum oxide tunnel barriers, inside a tunnel
junction with a nano-scale area small enough to allow individual
particles to be contacted.  (See device schematic, Fig.~1(a).)
The devices without a gate electrode are fabricated
\linebreak
\begin{figure}
\vspace{-0.8 cm}
\begin{center}
\leavevmode
\epsfxsize=7.2 cm
\epsfbox{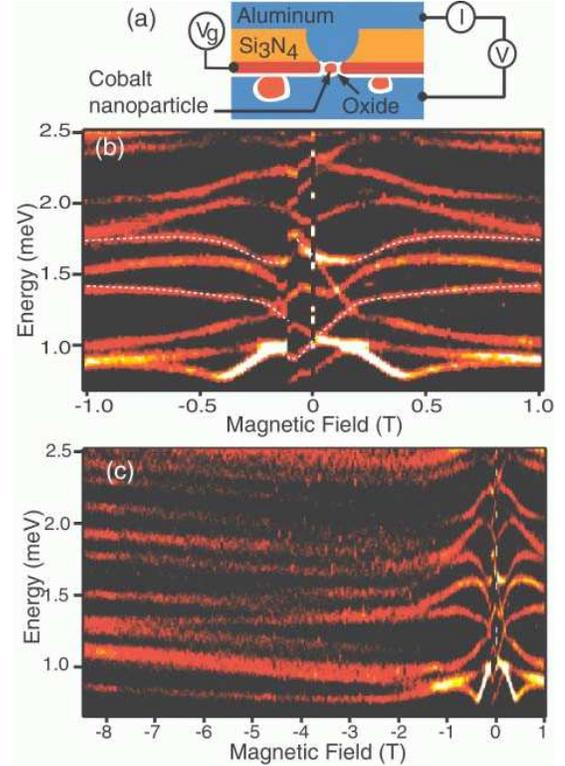}
\end{center}
\caption{ \label{figure1}
(a) Cross-sectional device
schematic. (b) and (c) Colorscale plots of $dI/dV$ for tunneling 
resonances in a cobalt nanoparticle. The field is varied from
positive to negative values, and $dI/dV$ vs.\ $V$ is measured at each 
step. The maximum conductance is
$3\times10^{-9} \mathrm \Omega^{-1}$.}
\end{figure}
\noindent
using a procedure
described previously \cite{gueron}, with the new innovation that by
using 1 nm of deposited aluminum oxide to form the tunnel barrier on
the cobalt particle we produce junctions with reduced charge noise.
The nanoparticles are made by depositing 0.5~nm of Co at room
temperature, which makes particles in the range of 1$-$4 nm diameter
\cite{gueron}.
The gated device is made by forming a hole about 10~nm in diameter in
a suspended silicon-nitride membrane, depositing 18.5~nm of Al to make the
gate electrode on the lower side of the device as shown in Fig.~1(a),
and then isolating this gate by anodizing in an oxygen plasma to 3.5 V
bias and depositing 8.5 nm of $\mathrm{SiO}_x$.
The rest of the fabrication proceeds in the same way as for the
non-gated devices.

When cooled to dilution refrigerator temperatures, the tunneling
conductance $dI/dV$ as a function of source-drain voltage $V$
consists of individual peaks, associated with transitions between
discrete electronic states in the nanoparticle \cite{gueron}.  The
$V$-spacings of resonances can be converted to energy, $\Delta
E=e\Delta V C_2/(C_1+C_2)$, in this way correcting for capacitive
division across the two junctions.  The capacitance ratio can be
determined by comparing peak positions at positive and negative $V$
\cite{ralph1,capacitances}.  In Fig.~1(b,c), we plot the energies
of tunneling resonances for a Co nanoparticle in a non-gated device
as a function of magnetic field, $H$.  As $H$ is swept from positive
values toward zero, the levels first undergo significant continuous
shifts.  The discontinuity near $H\!=\!0$ is an artifact of the
Al leads going superconducting and then being driven normal by
a negative field.  This causes the resonance energies to jump by 
$\Delta$, the superconducting gap \cite{ralph1}, but the energies of 
the states within the
nanoparticle evolve continuously.  Near $\mu_0 H_{\rm sw}=-0.120$~T,
all the levels exhibit another large discontinuity, which can be identified
with magnetic switching of the nanoparticle \cite{gueron}.  If $H$ is
swept from negative to positive values, the field value for this
transition is hysteretic, and the conductance spectrum is identical
to that shown, but reflected about $H$=0.
At larger values of $|H|$ (Fig.~1(c)), all the levels move to
increasing energy, with slopes that correspond to effective
$g$-factors ranging from 0.06 to 1.1.  The fact that the $g$-factors are
reduced below 2 indicates that the resonances are not purely spin-up
or spin-down, but are mixed by spin-orbit scattering \cite{salinas}.

	These data confirm in more detail several observations
made previously \cite{gueron}: the mean
energy spacing is about $0.2$~meV, smaller than estimated for independent
electrons within a particle 1$-$4 nm in diameter; the energies are 
coupled to the orientation of the
particle's magnetic moment; and resonances with
a given high-field slope can jump either up or down at $H_{\rm sw}$.
In addition, these data reveal several important new features.
(1) More resonances are resolved than in the previous work, and it
is clear that every level has a different $H$ dependence.   (2) When the
resonances approach close to each other as a function of $H$, they
cross, with no evidence of avoided-crossing behavior.
(3) The levels can exhibit complicated
non-monotonic variations as a function of $H$, for instance the
levels marked with dots in Fig.~1(b).

The new observations are incompatible with the simplest
model proposed previously for states in magnetic
nanoparticles \cite{gueron}, which
assumed that all the states in a
particle could be described by a single anisotropy-
\linebreak
\begin{figure}
\vspace{-1.2 cm}
\begin{center}
\leavevmode
\epsfxsize=8 cm
\epsfbox{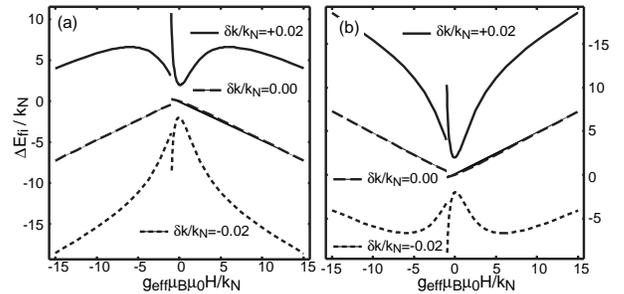}
\end{center}
\caption{
\label{figure2}
Tunneling energy $\Delta E_{fi}$ calculated using the semiclassical 
model discussed in the text, with $H$ at 45$^{\circ}$ from the easy 
axis, for various values of the anisotropy-energy difference $\delta 
k=k_{N+1}-k_{N}$.  In (a) the total spin is increasing; in (b) it is 
decreasing.
The curves are offset for clarity.}
\end{figure}
\noindent
energy
function.  This assumption has the consequence that all transitions
in which the spin component $\langle S_z\rangle$ increases should 
exhibit the same $H$
dependence, the $\langle S_z \rangle$-decreasing transitions should
show a second behavior, and neither dependence should exhibit the
complicated non-monotonicities below 1 Tesla evident in Fig.~1(b).
(The results of the older model are shown in the $\delta k=0$ curves
in Fig.~2.)  The model in \cite{gueron} assumed a uniaxial
anisotropy, but we have confirmed that the same behaviors hold for
more complicated forms which include terms of 2nd and 4th-order in
the spin-components \cite{thiaville}, as long as one assumes that the
same anisotropy function applies to all electronic states.

The fact that all the tunneling resonances undergo different energy
variations in the low-$H$ range where the magnetic moment is being
reoriented indicates directly that all the electronic states of the particle
cannot be described by the same anisotropy-energy function.
We have
explored whether such variations may also affect the form for the 
$H$-dependence
of the energies as described in point (3) above, and we find that 
they provide a
natural explanation for the complicated non-monotonic behavior as a
function of $H$.  We start by considering one single resonance 
associated with a transition between two states with $N$ and 
$N\!+\!1$ electrons.  We
extend the $N$-electron Hamiltonian
stated in \cite{gueron} in the simplest way to incorporate variations
in anisotropy energy:
\begin{equation}
\label{hamiltonian}
{\cal H}=-g_{\rm eff}\mu_{B}\mu_0\vec{H}\cdot\vec{S}
-k_{N} (\vec S \cdot \hat n)^2 /S_0 \, .
\end{equation}
Here $\hbar \vec{S}$ is the total spin with ground-state magnitude 
$\hbar S_0$ for $N$ electrons,
$\hat n$ is a unit vector in an
easy-axis direction, and the (uniaxial) anisotropy energy prefactor
$k_{N}$ is now allowed to vary between the $N$- and 
($N\!+\!1$)-electron states.  For simplicity,
we assume that the easy axis is the same for all states.
We have solved for the ground state energies for $N$ and $N+1$
electrons semiclassically as a function of $H$ by finding the spin 
orientation that gives the local minimum in Eq.~(1), assuming that 
$S_0$ does not vary with $H$
\cite{canali}, and then we calculate the form of the tunneling
transition energies as $E(N\!+\!1,H)-E(N,H)$.  The results are shown in
Fig.~2.  Assuming $S_0 \sim$ 1000 (appropriate to a 4 nm
Co particle), and an average value of $\langle k_{N}\rangle \approx$
0.01 meV in accord with the switching field \cite{kleff,correction},
fluctuations in $k_{N}$ of order 1-3\% are sufficient to explain both 
the size and form of the non-monotonicities.  Subsequent to this 
initial semiclassical analysis, similar conclusions were also reached 
in a more rigorous
quantum-mechanical picture \cite{kleff}.

Although fluctuations in the properties of eigenstates are
not often considered in the context of ferromagnets, they are not
surprising. In
non-magnetic particles, the $g$-factors for
Zeeman splitting fluctuate \cite{salinas}, and the statistics of these
fluctuations have been investigated in random-matrix treatments of
the spin-orbit interaction \cite{matveev,brouwer}.  Anisotropy in
magnetic particles also arises from spin-orbit interactions.  An
additional consequence of anisotropy-energy fluctuations should be that the
value of $H_{\rm sw}$ will vary by 1-3\% depending on the occupation of
excited electronic states.  This has yet to be investigated.


We now turn to data from a gated tunneling device, with the primary
motivation being
to resolve the question \cite{gueron} of whether the
larger-than-anticipated density of tunneling resonances observed at
low energy (noted above) is due to inelastic emission of spin-waves
during tunneling \cite{zhang} or due to non-equilibrium
effects \cite{kleff,agam}.  The idea of the
non-equilibrium mechanism is that the energy of tunneling transitions
can be described as the energy differences of states in the
nanoparticle with $N$ and $N\pm 1$ electrons:
$\Delta E^{\pm}_{fi} \equiv E_f^{N\pm 1} - E_i^N$.
If under conditions of current flow, $ E_i^N$ can assume an ensemble
of different values, for instance due to low-energy spin excitations
or electron-hole excitations
within the particle, then the number of observed tunneling resonances
can increase above the number that originate only from a single
equilibrium ground state.
We can test this with a gate voltage because the energy of a
tunneling transition can be tuned from high values down close
to zero where tunneling can be initiated by
small $V$.  For sufficiently small $V$, the tunneling electrons may
have insufficient energy to excite non-equilibrium states.
Therefore, a test of whether tunneling resonances are
associated with non-equilibrium initial states is whether some
transitions disappear when they are tuned to small $V$.  This
disappearance is exactly what is observed for the lowest-energy
transitions at $H$=0 (Fig.~3(a)), when the electrodes are
superconducting.

	Related non-equilibrium effects have been observed
in non-magnetic particles, but the consequences are much less
dramatic.  In Al particles, the energy shifts due to non-equilibrium
are small, resulting only in a fine-structure about the energy of
equilibrium transitions \cite{agam,deshmukh}.  We have not observed
well-resolved transitions in non-magnetic particles to disappear
completely as $V_g$ is varied, whereas at least the first 5
\linebreak
\begin{figure}
\vspace{-0.8cm}
\begin{center}
\leavevmode
\epsfxsize=7 cm
\epsfbox{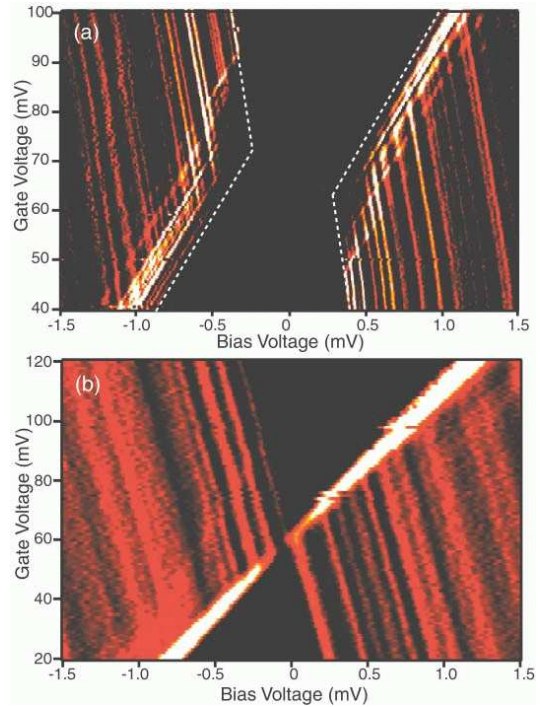}
\end{center}
\caption{
\label{figure3}
Colorscale conductance plots of a gated Co nanoparticle as a function
of $V_g$ and $V$. (a)  Superconducting leads ($H$=0).  (b) 
Normal-state leads (0.07 Tesla). The dashed lines in (a) indicate the 
expected evolution of the threshold
peaks if the resonances did not involve non-equilibrium processes. 
The voltage resolution is better at $H$=0 because of the 
superconducting singularity in the electrode density of states.  The 
electron $T\approx$ 90 mK.}
\end{figure}
\noindent
lowest-energy states in the Co particle lose conductance.
Non-equilibrium effects therefore appear to be much stronger in Co,
perhaps due to larger fluctuations in electron-electron interactions
and/or a large multiplicity of low-energy collective
spin excitations (in addition to the particle-hole excitations
that were considered previously \cite{agam}).
Since the relaxation rate of non-equilibrium excitations must be
slower than the tunneling rate for the spectrum to be
affected, the relaxation rate is of the order of or slower than
$\sim$ 1~MHz.

The presence of level crossings (instead of avoided crossings),
noted in Fig.~1 provides independent new evidence
supporting the non-equilibrium scenario.  In non-magnetic particles, 
when spin-orbit
scattering reduces the large-$H$ g-factors to
less than 1.7, tunneling transitions originating from the same initial
state exhibit clear level repulsion \cite{salinas}.  In contrast,
under non-equilibrium conditions, tunneling resonances occurring at
similar values of $V$ can result from different pairs of eigenstates
($E_i^N, E_f^{N\pm 1}$), none of which are nearly degenerate, so an
avoided crossing would not be expected.

Despite these two lines of evidence for the importance of
non-equilibrium transitions, our observations are not
\linebreak
\begin{figure}
\vspace{-1.0 cm}
\begin{center}
\leavevmode
\epsfxsize=7 cm
\epsfbox{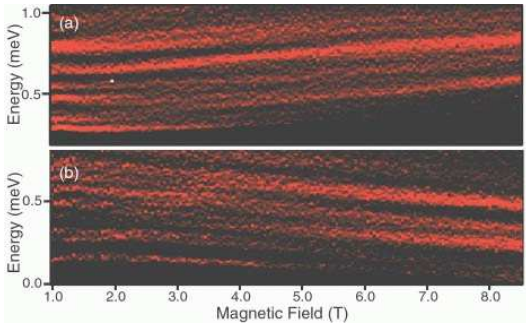}
\end{center}
\caption{
\label{figure4}
High-field conductance spectra for the sample of Fig.~3 at (a) 
$V_g=20$~mV and (b) $V_g=125$~mV.}
\end{figure}
\noindent
in full
agreement with the simplest phenomenological scenario that includes
non-equilibrium effects \cite{kleff}.  Ref. \cite{kleff} proposes
that non-equilibrium
spin-accumulation may occur by a sequence of transitions up a ladder
of energy states having different total spin $S$, with the sequence
terminating when the next step up the ladder requires more energy
than that provided by the source-drain bias, $eV$.  Within this
scenario, a fraction of both the low-energy and high-energy tunneling
resonances should disappear as $V_g$ tunes the transitions to lower
voltages.  In Fig.~3(a), only the lowest-energy transitions
experience a loss of tunneling amplitude.  We suspect that a more
complicated set of states than was used
in \cite{kleff} may need to be considered in characterizing
non-equilibrium processes on our magnetic nanoparticle.
For instance, single-electron quasiparticle excitations and also 
ladders of states with the same value of total spin $S$ but different 
values of $S_z$ may be involved.

Another
unanticipated observation is that the disappearance of tunneling
resonances as a function of $V_g$ occurs only when the electrodes are
superconducting, and not when they are normal (Fig.~3(b)).  Scans at
0.07, 1, 3, 5, and 8.5 T are all qualitatively similar. Perhaps, because of the
singularity in the density of states for superconducting electrodes,
the ensemble of non-equilibrium states may be populated differently
by tunneling from normal and superconducting electrodes, with less 
non-equilibrium at low $V$ in the superconducting case.  However, 
this process has
not yet been modeled in detail.

One more  simple benefit of having a gate electrode is that it
allows a determination of whether the tunneling resonances correspond
predominantly to the motion of majority or minority electrons.
Consider the data shown in Figures 3 and 4.  At $V_g=20$ mV the
low-energy tunneling thresholds correspond to the addition of an
electron to the nanoparticle ($N\!\rightarrow\!N\!+\!1$), and at large
magnetic fields  (Fig.~4(a)) the tunneling energies increase, meaning
that the tunneling states have lower $\langle S_z\rangle$.  It
follows that the transitions correspond predominantly to
minority-electron tunneling.  Similarly, at $V_g=125$ mV (Fig.~4(b)), the
tunneling thresholds are $N\!+\!1\! \rightarrow \!N$ processes in which
the spin increases, so a minority electron is tunneling off the
nanoparticle.  These results from one sample are in agreement with
predictions \cite{canali,kleff} that the total
spin should almost always decrease when electrons are added, due to 
both a large minority density of states and strong exchange 
interactions.  Of
course, because of spin-orbit coupling,
the transitions are not purely of minority-electron character.

In summary, measurements of the $H$-dependence of the 
tunneling-resonance energies in cobalt nanoparticles exhibit 
level-to-level differences and non-monotonic variations that can both 
be explained by variations of 1-3\% in magnetic anisotropy energy 
between different states.  The observation of level crossings and the 
disappearance of resonances as a function of $V_g$ indicate that the 
majority of low-energy resonances are associated with tunneling 
transitions from non-equilibrium initial states.  The strength of 
non-equilibrium effects appears to depend on whether the electrodes 
are normal or superconducting.

Acknowledgements: NSF DMR-0071631, DAAD19-01-1-0541, the Packard Foundation,
the Cornell Nanofabrication Facility, and the DFG-Program ``Semiconductor
and  Metallic Clusters".

\end{multicols}


\begin{thebibliography}{99}
\bibitem{correction}  The estimate $\langle k_{N}\rangle \sim$ 0.1
meV in \cite{gueron} based on the size of energy-level jumps at
$H_{\rm sw}$ is inaccurate because it neglects effects of
$k_N$-fluctuations \cite{kleff}.
\end{thebibliography}
\end{document}